%% file: preprint_JSM_2019.tex
\documentclass[12pt]{iopart}

\usepackage{graphicx}
\usepackage{amsfonts}
\usepackage{psfrag}

\newcommand{\pc}{p_{\scriptstyle \rm c}}
\newcommand{\pcI}{p_{\scriptstyle \rm c}^{\scriptstyle \rm I}}
\newcommand{\pcII}{p_{\scriptstyle \rm c}^{\scriptstyle \rm II}}

\begin{document}

\title{Selective Bootstrap Percolation}

\author{Mauro Sellitto}

\address{Dipartimento di Ingegneria, Universit\`a degli Studi della
  Campania ``Luigi Vanvitelli'', Via Roma 29, 81031 Aversa,
  Italy. \\ The Abdus Salam International Centre for Theoretical
  Physics, Strada Costiera~11, 34151 Trieste, Italy.  }
\ead{mauro.sellitto@unicampania.it, mauro.sellitto@gmail.com}

\begin{abstract} 
A new class of bootstrap percolation models in which particle culling
occurs only for certain numbers of nearest neighbours is introduced
and studied on a Bethe lattice. Upon increasing the density of initial
configuration they undergo multiple hybrid (or mixed-order) phase
transitions, showing that such intriguing phase behaviours may also
appear in fully homogeneous situations/environments, provided that
culling is {\em selective} rather than {\em cumulative}.  The idea
immediately extends to facilitation dynamics, suggesting a simple way
to construct one-component models of multiple glasses and glass-glass
transitions as well as more general coarse-grained models of complex
cooperative dynamics.
\end{abstract}

\maketitle

\date{\today}

\section*{Introduction}

Bootstrap percolation (BP) is a primary example of a statistical
mechanics model in which a discontinuous phase transition is
accompanied by critical fluctuations~\cite{Chalupa}. Interest in this
unsual hybrid (or mixed-order) phase behaviour has grown in the last
few years while it has been discovered in a variety of quite unrelated
systems~\cite{BP_rev,Bar}. Prominent examples of interest here are
certain models of complex networks~\cite{Branco,Doro,Cellai}, and
especially of glassy systems, where the possibility of multiple hybrid
phase transitions and higher-order
singularities~\cite{SeDeCaAr,Se2012,PaSe}, similar to those first
discovered in the mode-coupling theory~\cite{Dawson,Goetze}, was
inferred through the close connection of BP with facilitation glassy
dynamics~\cite{RiSo,SeBiTo}. These results were generally obtained at
the price of introducing some form of inhomogeneity in the problem,
such as a fluctuating connectivity in the underlying lattice structure
or multiple culling
thresholds~\cite{Branco,Doro,Cellai,SeDeCaAr,Se2012,PaSe}, according
to which a particle is culled if a cumulative constraint on the number
of its nearest neighbours is satisfied in the form of inequality ({\em
  less than} or, equivalently, {\em greater than}). In this paper, we
show that this inhomogeneity is not a necessary ingredient and one may
find multiple hybrid phase transitions in fully homogeneous
environments provided that particle culling is selective, i.e., it
does involve only certain specific numbers of nearest neighbours.

There are two main general motivations behind this work.  The first
stems from the difficulty of finding one-component schematic models of
multiple glasses which are sufficiently simple to allow for large
scale numerical simulations or even analytical solutions of their
equilibrium and aging dynamics.  Multiple hybrid phase transitions in
such systems, generally arise from the presence of two (or more)
microscopic different length scales (typically a repulsive hard-core
plus a short-range attractive or even repulsive
potential)~\cite{Sciortino,Pham,Voigtmann,Gnan,Maimbourg}. It is the
competition between these length scales that leads, as the
thermodynamic control parameters are changed, to multiple phases
dominated by qualitatively different particle packing mechanisms on
different length scales.

The second motivation is that the idea put forward here may be useful,
more broadly, for understanding certain forms of complex dynamics in
which cooperativity does not necessarily involves a cumulative effect
but rather a selective one. It is well known, for example, that many
soft matter systems dominated by hydrophobic
interactions~\cite{Nelson}, most notably proteins, loose their
biological functionality when thermodynamics variables (temperature,
pressure, pH concentration) lie outside a relatively narrow range
around their ambient values, leading to unusual {\em reentrant}
Tammann phase diagrams~\cite{Greer,StDe,ScSh,Plazanet}. Another
example of interest is offered by the evolutionary dynamics of
ecological systems whose stationary states are typically neither
scarcely populated nor overcrowded, as epitomized by the celebrated
Game of Life cellular automaton, whose dynamical rules are strongly
reminiscent of selective kinetic constraints, albeit of irreversible
type~\cite{Schulman,BaReRu}.

\section*{Boostrap percolation}

BP is generally concerned with the statistical properties of particle
clusters generated on a lattice according to the following
procedure. The sites of a lattice with coordination number $k+1$ are
first occupied with probability $p$ as in ordinary percolation. Then,
particles with at least $f$ vacant neighbours are iteratively removed
one by one, until a static configuration is reached\footnote{Note that
  our definition of the bootstrap percolation rule is equivalent, on
  the regular lattice considered here, to the one traditionally used
  in the literature, in which a particle is culled if it has less than
  $m=k-f+2$ neighbouring particles.  This is done with the purpose of
  making a more direct connection to works on facilitated glassy
  dynamics in which a spin can flip only when it has at least $f$
  nearest neighbouring down-spin state.}. Depending on $p$, the final
configuration can be either empty or an $f$-cluster built up by
particles with less than $f$ vacant neighbours. The passage between
these two states may have intriguing critical features. On locally
tree-like lattices with fixed branching ratio $k$, the case we
consider here, an infinite $f$-cluster can only exist for $f>1$ and
the BP problem can be solved exactly~\cite{Chalupa}.  The residual
fraction, $\Phi$, of occupied site represents the system order
parameter and can be written as:
\begin{eqnarray}
    \Phi = p \ \sum_{i \, \in \, {\mathcal S}} {k+1 \choose i} B^i
    (1-B)^{k+1-i} ,
\label{eq.Phi}
\end{eqnarray}
where ${\mathcal S}= \{ 0,\,1,\, \cdots,\,f-1\}$ is the set of numbers
of nearest neighbours that prevent particle culling, and $B$ is the
probability that a site is {\em not} connected to the infinite
$f$-cluster through a nearest neighbour, which obeys the
self-consistent polynomial equation:
\begin{equation}
  1 - B = p \ \sum_{i \, \in \, {\mathcal S}} {k \choose i} B^i
  (1-B)^{k-i} .
\label{eq.B}
\end{equation}
The critical properties of BP transitions are easily established by
Taylor expanding the right member side of Eq.~\ref{eq.B} in powers of
$1-B$. The presence of a linear term for $f=k$ implies a continuous
phase transition located at at $\pc=1/k$. The incipient spanning
$f$-cluster in this case is fractal, as the model reduces to the usual
random percolation. When $ 1 < f < k$, instead, the absence of a
linear term in $1-B$ implies that $B$, and therefore $\Phi$, must
change abruptly: the phase transition is discontinuous and the
structure of $f$-cluster is compact. However, since $\Phi$ has a
square root singularity at the transition the diverging susceptibility
$d\Phi/dp$ implies critical fluctuations. The underlying geometrical
nature of this behaviour is the diverging mean size of {\it corona}
clusters (i.e., clusters in which every particles has exactly $f-1$
vacant neighbours) and their structural fragility (corona
collapse occurs, as in domino-like games, by a single particle
culling)~\cite{Doro}.

\section*{Selective Bootstrap Percolation}

In the more general selective bootstrap percolation (SBP) we now
introduce, ${\mathcal S}$ is an {\em arbitrary} subset of non-negative
integers less than $f$. The critical properties of the phase
transition first encountered by increasing $p$ is determined by the
lowest order terms in $1-B$ of the equation $ {\mathcal Q}=0$, with:
\begin{equation} {\mathcal Q}(B) = 1 - B - p \ \sum_{i \, \in \,
    {\mathcal S} } {k \choose i} B^{i} (1-B)^{k-i} ,
\label{eq.Q}
\end{equation}
which depends only on the maximum value of ${\mathcal S}$, that is
$\mathsf{max} \,{\mathcal S} \equiv f-1$. Therefore we fully recover
the two types of phase behaviour previously discussed for BP: a
continuous transition located at $\pc=1/k$ when $\mathsf{max} \,
\mathcal{S}=k-1$, and a discontinuous hybrid transition for $2<
\mathsf{max} \,{\mathcal S}<k-1$.  From a purely mathematical point of
view, the main novelty of SBP is that, when gaps in the sequence of
non-negative integers less than $f$ are generally allowed, the absence
of some powers of $B$ in the polynomial ${\mathcal Q}(B)$ will
possibly lead to the appearance of extra real roots and, consequently,
to multiple hybrid phase transitions.
From a more geometrical/physical point of view, the selective kinetic
constraint allows for the formation of different corona clusters,
whose competition and stability lead to the existence of multiple
hybrid behaviour, as the density of the initial particle configuration
is changed.
Moreover, by simply increasing the lattice connectivity one can
further augment the number of real roots (up to $k$, virtually, by the
fundamental theorem of algebra).  Therefore, we can generally conclude
that SBP displays a phase behaviour richer than BP, 
even though the basic geometric mechanism of the hybrid behaviour in
SBP remains essentially the same as that of BP.

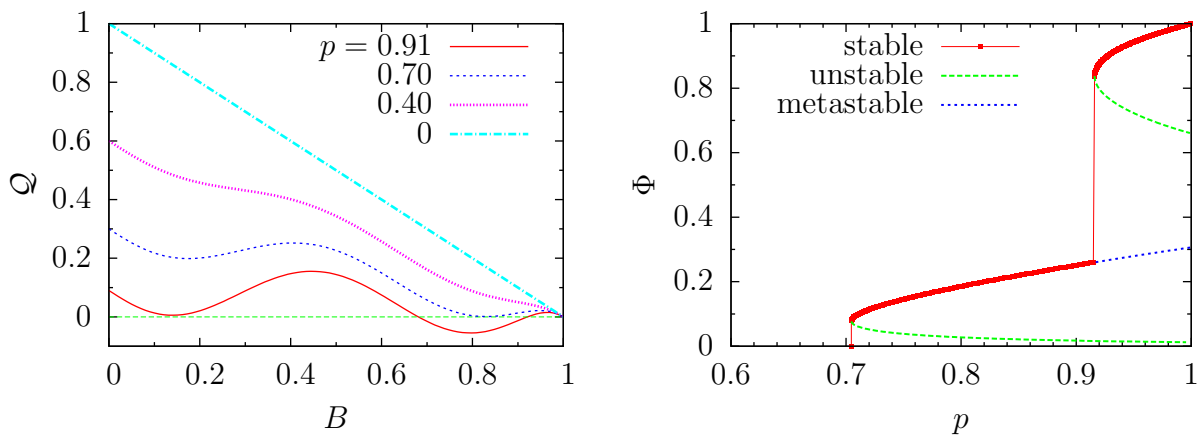
\begin{figure}[htbp]
\input{pot_c8_disc} \input{phi_c8_disc_v2}
\caption{Effective potential ${\mathcal Q}$ and order parameter $\Phi$
  in a Bethe lattice with connectivity $z=8$ and selective constraint
  ${\mathcal S} = \{0 ,\,1,\,2,\,5\}$.}
\label{fig.disc_c8}
\end{figure}

Before illustrating explicitly these results in two cases, it is
important to emphasize that the correct determination of the phase
diagram and the evolution of the order parameter $\Phi$ in the
presence of multiple phase transitions, requires that a stability
criterion for the selection of physical stable solution is properly
established.
This is simply done by observing that: $i)$ the particle fraction
$\rho_t$ populating the lattice at time $t$, while particle culling is
carried out, is a non-increasing function of $t$, and $ii)$ the
asymptotic density $\rho_{\infty}=\Phi$ is a monotonic increasing
function of the initial particle fraction $\rho_0=p$. This implies
that the stable physical solution corresponds to the value of the
residual fraction $\Phi$ or, equivalently, $B$, which is first
encountered during culling dynamics, namely to the largest root of
Eq.~(\ref{eq.Q}).

The first case we consider is a Bethe lattice with branching ratio
$k=7$ and ${\mathcal S} = \{0 ,\,1,\,2,\,5\}$ which gives the
self-consistent equation:
\begin{equation}
  1-B=p \left[ (1-B)^7 + 7 (1-B)^6 B + 21 (1-B)^5 B^2 + 21 (1-B)^2 B^5
    \right] .
\label{eq.Q_c8_disc}
\end{equation}
As shown in Fig.~\ref{fig.disc_c8} the potential ${\mathcal Q}$ has
two real roots.  Accordingly, the order parameter $\Phi$ undergoes two
jumps with a square-root singularity (the first located at $\pcI
\simeq 0.7046...$ and the second at $\pcII \simeq 0.9149..$) implying
critical fluctuations and therefore a double hybrid phase transition,
as expected (since $\mathsf{max} \, \mathcal{S}=5<k-1=6$).

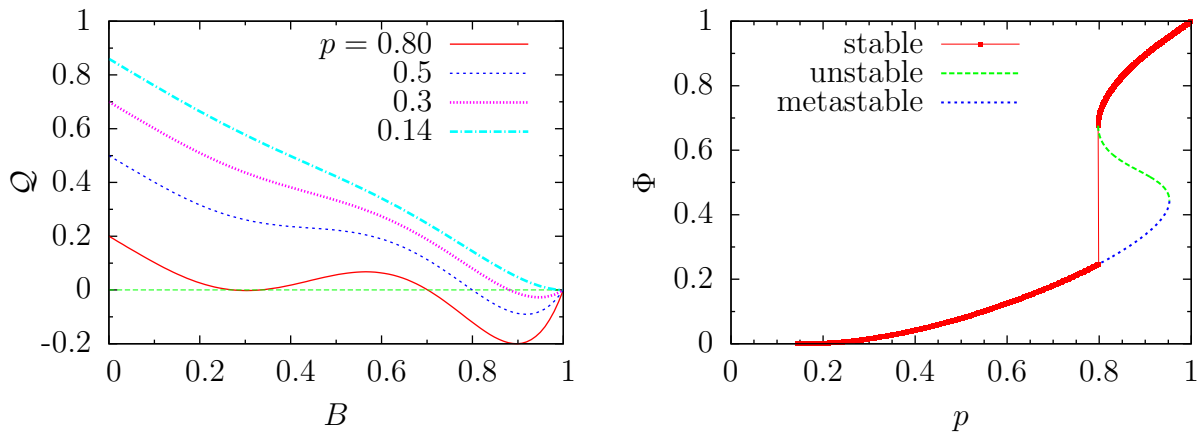
\begin{figure}[htbp]
\input{pot_c8_cont} \input{phi_c8_cont_v2}
\caption{Effective potential ${\mathcal Q}$ and order
  parameter $\Phi$ in a Bethe lattice with branching ratio $k=7$ and
  selective constraint ${\mathsf S} = \{0 ,\,1,\,2,\,3,\,6\}$.}
\label{fig.cont_c8}
\end{figure}

Next we consider the set ${\mathcal S} = \{0 ,\,1,\,2,\,3,\,6\}$ for a
Bethe lattice with $k=7$. This gives the self-consistent equation:
\begin{equation}
  1-B=p \left[ 
1 - B^7
- 21 (1-B)^2 B^5 -35  (1-B)^3 B^4 
    \right] .
\label{eq.Q_c8_disc}
\end{equation}
As shown in Fig.~\ref{fig.cont_c8} the potential ${\mathcal Q}$ has
again two real roots but now the order parameter $\Phi$ undergoes
first a continuous phase transition at $\pcI = 1/7 \simeq 0.1437...$
(as expected, since $\mathsf{max} \, \mathcal{S}=6=k-1$), and then a
hybrid phase transition located at $\pcII \simeq 0.7986..$.  
To highlight the {\em fold} (or, equivalently, {\em saddle-node})
bifurcation of the discontinuous transition we show, in the right
panel of both Figs.~\ref{fig.disc_c8} and~\ref{fig.cont_c8}, the
unstable and metastable solutions. The former corresponds to the order
parameter decreasing with $p$, which clearly violates the condition
$ii)$ of the stability criterion of physical solution previously
discussed. The latter, also violates that criterion, nevertheless, in
facilitation dynamics closely related to BSP, the observation of
metastable solutions in out of equilibrium states, e.g., during a
sufficiently slow annealing and for suitable initial conditions, is
possible.

Finally, it should be noticed that although the number of jumps in the
order parameter can be increased by introducing additional gaps in the
sequence of integers of ${\mathcal S}$ and making $k$ larger, the
absence of a continuously tunable parameter makes it difficult to find
${\mathsf A}_{\ell}$ bifurcation singularities (at which the order
parameter critical exponent is $\beta=1/\ell$).  The appearance of
such higher-order critical points, corresponding to an
$\ell$-degeneracy ($\ell \ge 2$) of the largest root of ${\mathcal
  Q}$, that is:
\begin{equation}
\frac{d^n {\mathcal Q}}{dB^n} = 0 \,, \qquad n=0, \cdots, \ell-1,
\label{eq.dQ}
\end{equation}
with the $\ell$th derivative being nonzero, though a priori not
impossible for suitable values of $k$ and ${\mathcal S}$, must be
considered rather accidental.

\section*{Conclusions}

To conclude, we have introduced a simple and rather straigthforward
modification of BP featuring multiple hybrid phase transitions on a
random regular graph with fixed connectivity (i.e., a Bethe lattice),
in a fully homogeneous environment.
We expect this extension of the BP framework to be relevant to the
field of complex networks and other forms of cooperative dynamical
problems.
In particular, the connection between BP and facilitation dynamics
immediately suggests how to construct cooperative models with
selective facilitation: a non-interacting spin system in which a spin
can flip only when the number of its down-spin nearest neighbours
belongs to the set ${\mathcal S}$, will have a relaxation dynamics
with multiple phases corresponding to the fixed points of SBP
Eq.~(\ref{eq.B}). A similar procedure also applies to kinetically
constrained particle models.  By exploiting this connection one could
address the relaxation dynamics of multiple glasses, a problem that
seems at the moment hardly affordable in mean-field disordered
$p+q$-spin models, as the one-step replica symmetry breaking solution
seems to be unstable in this type of systems~\cite{Krako,CrLe}.  Also,
it would be interesting to investigate, e.g., with the $M$-layer
construction~\cite{Rizzo}, how multiple glass transitions are possibly
avoided in physical dimension~\cite{RiVo,PaRi}. This might shed some
light on the nature of ergodicity breaking in multiple glasses
observed experimentally and numerically in short-range attractive
colloids and related systems~\cite{Pham,Voigtmann,Gnan}.



\end{document}

%% file: pot_c8_disc.tex
\begingroup
  \makeatletter
  \providecommand\color[2][]{%
    \GenericError{(gnuplot) \space\space\space\@spaces}{%
      Package color not loaded in conjunction with
      terminal option `colourtext'%
    }{See the gnuplot documentation for explanation.%
    }{Either use 'blacktext' in gnuplot or load the package
      color.sty in LaTeX.}%
    \renewcommand\color[2][]{}%
  }%
  \providecommand\includegraphics[2][]{%
    \GenericError{(gnuplot) \space\space\space\@spaces}{%
      Package graphicx or graphics not loaded%
    }{See the gnuplot documentation for explanation.%
    }{The gnuplot epslatex terminal needs graphicx.sty or graphics.sty.}%
    \renewcommand\includegraphics[2][]{}%
  }%
  \providecommand\rotatebox[2]{#2}%
  \@ifundefined{ifGPcolor}{%
    \newif\ifGPcolor
    \GPcolortrue
  }{}%
  \@ifundefined{ifGPblacktext}{%
    \newif\ifGPblacktext
    \GPblacktexttrue
  }{}%
  \let\gplgaddtomacro\g@addto@macro
  \gdef\gplbacktext{}%
  \gdef\gplfronttext{}%
  \makeatother
  \ifGPblacktext
    \def\colorrgb#1{}%
    \def\colorgray#1{}%
  \else
    \ifGPcolor
      \def\colorrgb#1{\color[rgb]{#1}}%
      \def\colorgray#1{\color[gray]{#1}}%
      \expandafter\def\csname LTw\endcsname{\color{white}}%
      \expandafter\def\csname LTb\endcsname{\color{black}}%
      \expandafter\def\csname LTa\endcsname{\color{black}}%
      \expandafter\def\csname LT0\endcsname{\color[rgb]{1,0,0}}%
      \expandafter\def\csname LT1\endcsname{\color[rgb]{0,1,0}}%
      \expandafter\def\csname LT2\endcsname{\color[rgb]{0,0,1}}%
      \expandafter\def\csname LT3\endcsname{\color[rgb]{1,0,1}}%
      \expandafter\def\csname LT4\endcsname{\color[rgb]{0,1,1}}%
      \expandafter\def\csname LT5\endcsname{\color[rgb]{1,1,0}}%
      \expandafter\def\csname LT6\endcsname{\color[rgb]{0,0,0}}%
      \expandafter\def\csname LT7\endcsname{\color[rgb]{1,0.3,0}}%
      \expandafter\def\csname LT8\endcsname{\color[rgb]{0.5,0.5,0.5}}%
    \else
      \def\colorrgb#1{\color{black}}%
      \def\colorgray#1{\color[gray]{#1}}%
      \expandafter\def\csname LTw\endcsname{\color{white}}%
      \expandafter\def\csname LTb\endcsname{\color{black}}%
      \expandafter\def\csname LTa\endcsname{\color{black}}%
      \expandafter\def\csname LT0\endcsname{\color{black}}%
      \expandafter\def\csname LT1\endcsname{\color{black}}%
      \expandafter\def\csname LT2\endcsname{\color{black}}%
      \expandafter\def\csname LT3\endcsname{\color{black}}%
      \expandafter\def\csname LT4\endcsname{\color{black}}%
      \expandafter\def\csname LT5\endcsname{\color{black}}%
      \expandafter\def\csname LT6\endcsname{\color{black}}%
      \expandafter\def\csname LT7\endcsname{\color{black}}%
      \expandafter\def\csname LT8\endcsname{\color{black}}%
    \fi
  \fi
  \setlength{\unitlength}{0.0500bp}%
  \begin{picture}(4680.00,3402.00)%
    \gplgaddtomacro\gplbacktext{%
      \csname LTb\endcsname%
      \put(726,925){\makebox(0,0)[r]{\strut{} 0}}%
      \put(726,1368){\makebox(0,0)[r]{\strut{} 0.2}}%
      \put(726,1810){\makebox(0,0)[r]{\strut{} 0.4}}%
      \put(726,2252){\makebox(0,0)[r]{\strut{} 0.6}}%
      \put(726,2695){\makebox(0,0)[r]{\strut{} 0.8}}%
      \put(726,3137){\makebox(0,0)[r]{\strut{} 1}}%
      \put(858,484){\makebox(0,0){\strut{} 0}}%
      \put(1543,484){\makebox(0,0){\strut{} 0.2}}%
      \put(2228,484){\makebox(0,0){\strut{} 0.4}}%
      \put(2913,484){\makebox(0,0){\strut{} 0.6}}%
      \put(3598,484){\makebox(0,0){\strut{} 0.8}}%
      \put(4283,484){\makebox(0,0){\strut{} 1}}%
      \put(220,1920){\rotatebox{-270}{\makebox(0,0){\strut{}${\mathcal Q}$}}}%
      \put(2570,154){\makebox(0,0){\strut{}$B$}}%
    }%
    \gplgaddtomacro\gplfronttext{%
      \csname LTb\endcsname%
      \put(3296,2964){\makebox(0,0)[r]{\strut{}$p=0.91$}}%
      \csname LTb\endcsname%
      \put(3296,2744){\makebox(0,0)[r]{\strut{}$0.70$}}%
      \csname LTb\endcsname%
      \put(3296,2524){\makebox(0,0)[r]{\strut{}$0.40$}}%
      \csname LTb\endcsname%
      \put(3296,2304){\makebox(0,0)[r]{\strut{}$0$}}%
    }%
    \gplbacktext
    \put(0,0){\includegraphics{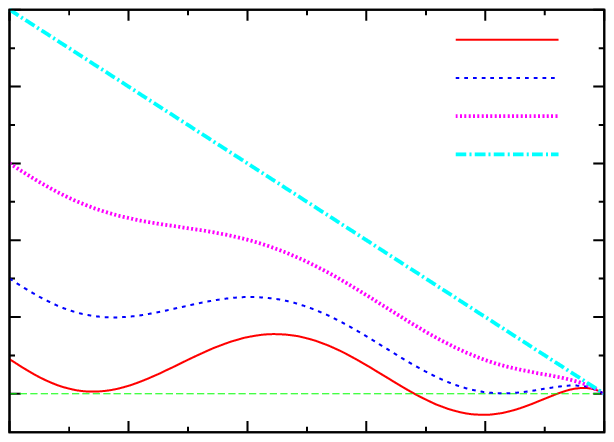}}%
    \gplfronttext
  \end{picture}%
\endgroup

%% file: phi_c8_disc_v2.tex
\begingroup
  \makeatletter
  \providecommand\color[2][]{%
    \GenericError{(gnuplot) \space\space\space\@spaces}{%
      Package color not loaded in conjunction with
      terminal option `colourtext'%
    }{See the gnuplot documentation for explanation.%
    }{Either use 'blacktext' in gnuplot or load the package
      color.sty in LaTeX.}%
    \renewcommand\color[2][]{}%
  }%
  \providecommand\includegraphics[2][]{%
    \GenericError{(gnuplot) \space\space\space\@spaces}{%
      Package graphicx or graphics not loaded%
    }{See the gnuplot documentation for explanation.%
    }{The gnuplot epslatex terminal needs graphicx.sty or graphics.sty.}%
    \renewcommand\includegraphics[2][]{}%
  }%
  \providecommand\rotatebox[2]{#2}%
  \@ifundefined{ifGPcolor}{%
    \newif\ifGPcolor
    \GPcolortrue
  }{}%
  \@ifundefined{ifGPblacktext}{%
    \newif\ifGPblacktext
    \GPblacktexttrue
  }{}%
  \let\gplgaddtomacro\g@addto@macro
  \gdef\gplbacktext{}%
  \gdef\gplfronttext{}%
  \makeatother
  \ifGPblacktext
    \def\colorrgb#1{}%
    \def\colorgray#1{}%
  \else
    \ifGPcolor
      \def\colorrgb#1{\color[rgb]{#1}}%
      \def\colorgray#1{\color[gray]{#1}}%
      \expandafter\def\csname LTw\endcsname{\color{white}}%
      \expandafter\def\csname LTb\endcsname{\color{black}}%
      \expandafter\def\csname LTa\endcsname{\color{black}}%
      \expandafter\def\csname LT0\endcsname{\color[rgb]{1,0,0}}%
      \expandafter\def\csname LT1\endcsname{\color[rgb]{0,1,0}}%
      \expandafter\def\csname LT2\endcsname{\color[rgb]{0,0,1}}%
      \expandafter\def\csname LT3\endcsname{\color[rgb]{1,0,1}}%
      \expandafter\def\csname LT4\endcsname{\color[rgb]{0,1,1}}%
      \expandafter\def\csname LT5\endcsname{\color[rgb]{1,1,0}}%
      \expandafter\def\csname LT6\endcsname{\color[rgb]{0,0,0}}%
      \expandafter\def\csname LT7\endcsname{\color[rgb]{1,0.3,0}}%
      \expandafter\def\csname LT8\endcsname{\color[rgb]{0.5,0.5,0.5}}%
    \else
      \def\colorrgb#1{\color{black}}%
      \def\colorgray#1{\color[gray]{#1}}%
      \expandafter\def\csname LTw\endcsname{\color{white}}%
      \expandafter\def\csname LTb\endcsname{\color{black}}%
      \expandafter\def\csname LTa\endcsname{\color{black}}%
      \expandafter\def\csname LT0\endcsname{\color{black}}%
      \expandafter\def\csname LT1\endcsname{\color{black}}%
      \expandafter\def\csname LT2\endcsname{\color{black}}%
      \expandafter\def\csname LT3\endcsname{\color{black}}%
      \expandafter\def\csname LT4\endcsname{\color{black}}%
      \expandafter\def\csname LT5\endcsname{\color{black}}%
      \expandafter\def\csname LT6\endcsname{\color{black}}%
      \expandafter\def\csname LT7\endcsname{\color{black}}%
      \expandafter\def\csname LT8\endcsname{\color{black}}%
    \fi
  \fi
  \setlength{\unitlength}{0.0500bp}%
  \begin{picture}(4680.00,3402.00)%
    \gplgaddtomacro\gplbacktext{%
      \csname LTb\endcsname%
      \put(682,704){\makebox(0,0)[r]{\strut{}$0$}}%
      \put(682,1191){\makebox(0,0)[r]{\strut{}$0.2$}}%
      \put(682,1677){\makebox(0,0)[r]{\strut{}$0.4$}}%
      \put(682,2164){\makebox(0,0)[r]{\strut{}$0.6$}}%
      \put(682,2650){\makebox(0,0)[r]{\strut{}$0.8$}}%
      \put(682,3137){\makebox(0,0)[r]{\strut{}$1$}}%
      \put(814,484){\makebox(0,0){\strut{}$0.6$}}%
      \put(1681,484){\makebox(0,0){\strut{}$0.7$}}%
      \put(2548,484){\makebox(0,0){\strut{}$0.8$}}%
      \put(3416,484){\makebox(0,0){\strut{}$0.9$}}%
      \put(4283,484){\makebox(0,0){\strut{}$1$}}%
      \put(176,1920){\rotatebox{-270}{\makebox(0,0){\strut{}$\Phi$}}}%
      \put(2548,154){\makebox(0,0){\strut{}$p$}}%
    }%
    \gplgaddtomacro\gplfronttext{%
      \csname LTb\endcsname%
      \put(2266,2964){\makebox(0,0)[r]{\strut{}stable}}%
      \csname LTb\endcsname%
      \put(2266,2744){\makebox(0,0)[r]{\strut{}unstable}}%
      \csname LTb\endcsname%
      \put(2266,2524){\makebox(0,0)[r]{\strut{}metastable}}%
    }%
    \gplbacktext
    \put(0,0){\includegraphics{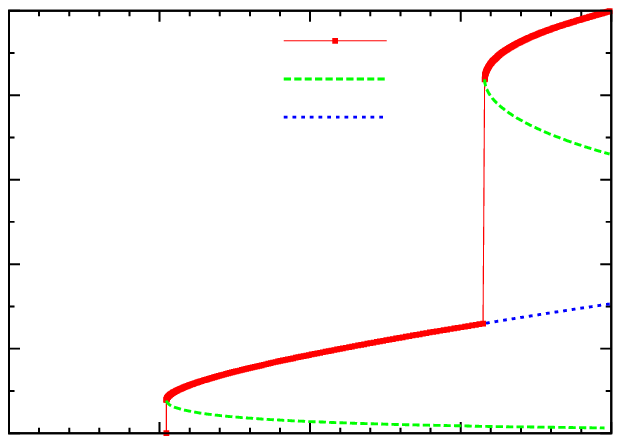}}%
    \gplfronttext
  \end{picture}%
\endgroup

%% file: pot_c8_cont.tex
\begingroup
  \makeatletter
  \providecommand\color[2][]{%
    \GenericError{(gnuplot) \space\space\space\@spaces}{%
      Package color not loaded in conjunction with
      terminal option `colourtext'%
    }{See the gnuplot documentation for explanation.%
    }{Either use 'blacktext' in gnuplot or load the package
      color.sty in LaTeX.}%
    \renewcommand\color[2][]{}%
  }%
  \providecommand\includegraphics[2][]{%
    \GenericError{(gnuplot) \space\space\space\@spaces}{%
      Package graphicx or graphics not loaded%
    }{See the gnuplot documentation for explanation.%
    }{The gnuplot epslatex terminal needs graphicx.sty or graphics.sty.}%
    \renewcommand\includegraphics[2][]{}%
  }%
  \providecommand\rotatebox[2]{#2}%
  \@ifundefined{ifGPcolor}{%
    \newif\ifGPcolor
    \GPcolortrue
  }{}%
  \@ifundefined{ifGPblacktext}{%
    \newif\ifGPblacktext
    \GPblacktexttrue
  }{}%
  \let\gplgaddtomacro\g@addto@macro
  \gdef\gplbacktext{}%
  \gdef\gplfronttext{}%
  \makeatother
  \ifGPblacktext
    \def\colorrgb#1{}%
    \def\colorgray#1{}%
  \else
    \ifGPcolor
      \def\colorrgb#1{\color[rgb]{#1}}%
      \def\colorgray#1{\color[gray]{#1}}%
      \expandafter\def\csname LTw\endcsname{\color{white}}%
      \expandafter\def\csname LTb\endcsname{\color{black}}%
      \expandafter\def\csname LTa\endcsname{\color{black}}%
      \expandafter\def\csname LT0\endcsname{\color[rgb]{1,0,0}}%
      \expandafter\def\csname LT1\endcsname{\color[rgb]{0,1,0}}%
      \expandafter\def\csname LT2\endcsname{\color[rgb]{0,0,1}}%
      \expandafter\def\csname LT3\endcsname{\color[rgb]{1,0,1}}%
      \expandafter\def\csname LT4\endcsname{\color[rgb]{0,1,1}}%
      \expandafter\def\csname LT5\endcsname{\color[rgb]{1,1,0}}%
      \expandafter\def\csname LT6\endcsname{\color[rgb]{0,0,0}}%
      \expandafter\def\csname LT7\endcsname{\color[rgb]{1,0.3,0}}%
      \expandafter\def\csname LT8\endcsname{\color[rgb]{0.5,0.5,0.5}}%
    \else
      \def\colorrgb#1{\color{black}}%
      \def\colorgray#1{\color[gray]{#1}}%
      \expandafter\def\csname LTw\endcsname{\color{white}}%
      \expandafter\def\csname LTb\endcsname{\color{black}}%
      \expandafter\def\csname LTa\endcsname{\color{black}}%
      \expandafter\def\csname LT0\endcsname{\color{black}}%
      \expandafter\def\csname LT1\endcsname{\color{black}}%
      \expandafter\def\csname LT2\endcsname{\color{black}}%
      \expandafter\def\csname LT3\endcsname{\color{black}}%
      \expandafter\def\csname LT4\endcsname{\color{black}}%
      \expandafter\def\csname LT5\endcsname{\color{black}}%
      \expandafter\def\csname LT6\endcsname{\color{black}}%
      \expandafter\def\csname LT7\endcsname{\color{black}}%
      \expandafter\def\csname LT8\endcsname{\color{black}}%
    \fi
  \fi
  \setlength{\unitlength}{0.0500bp}%
  \begin{picture}(4680.00,3402.00)%
    \gplgaddtomacro\gplbacktext{%
      \csname LTb\endcsname%
      \put(726,704){\makebox(0,0)[r]{\strut{}-0.2}}%
      \put(726,1110){\makebox(0,0)[r]{\strut{} 0}}%
      \put(726,1515){\makebox(0,0)[r]{\strut{} 0.2}}%
      \put(726,1921){\makebox(0,0)[r]{\strut{} 0.4}}%
      \put(726,2326){\makebox(0,0)[r]{\strut{} 0.6}}%
      \put(726,2732){\makebox(0,0)[r]{\strut{} 0.8}}%
      \put(726,3137){\makebox(0,0)[r]{\strut{} 1}}%
      \put(858,484){\makebox(0,0){\strut{} 0}}%
      \put(1543,484){\makebox(0,0){\strut{} 0.2}}%
      \put(2228,484){\makebox(0,0){\strut{} 0.4}}%
      \put(2913,484){\makebox(0,0){\strut{} 0.6}}%
      \put(3598,484){\makebox(0,0){\strut{} 0.8}}%
      \put(4283,484){\makebox(0,0){\strut{} 1}}%
      \put(220,1920){\rotatebox{-270}{\makebox(0,0){\strut{}${\mathcal Q}$}}}%
      \put(2570,154){\makebox(0,0){\strut{}$B$}}%
    }%
    \gplgaddtomacro\gplfronttext{%
      \csname LTb\endcsname%
      \put(3296,2964){\makebox(0,0)[r]{\strut{}$p=0.80$}}%
      \csname LTb\endcsname%
      \put(3296,2744){\makebox(0,0)[r]{\strut{}$0.5$}}%
      \csname LTb\endcsname%
      \put(3296,2524){\makebox(0,0)[r]{\strut{}$0.3$}}%
      \csname LTb\endcsname%
      \put(3296,2304){\makebox(0,0)[r]{\strut{}$0.14$}}%
    }%
    \gplbacktext
    \put(0,0){\includegraphics{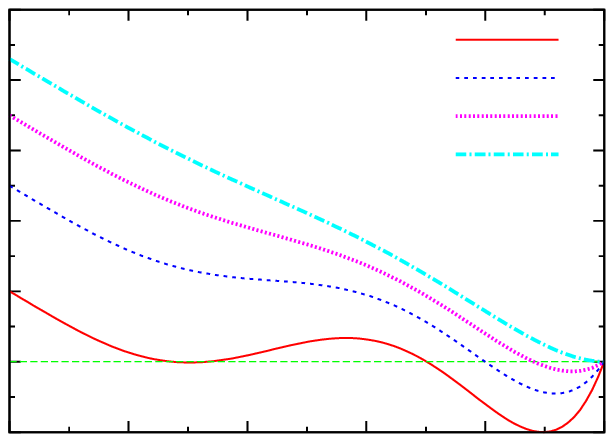}}%
    \gplfronttext
  \end{picture}%
\endgroup

%% file: phi_c8_cont_v2.tex
\begingroup
  \makeatletter
  \providecommand\color[2][]{%
    \GenericError{(gnuplot) \space\space\space\@spaces}{%
      Package color not loaded in conjunction with
      terminal option `colourtext'%
    }{See the gnuplot documentation for explanation.%
    }{Either use 'blacktext' in gnuplot or load the package
      color.sty in LaTeX.}%
    \renewcommand\color[2][]{}%
  }%
  \providecommand\includegraphics[2][]{%
    \GenericError{(gnuplot) \space\space\space\@spaces}{%
      Package graphicx or graphics not loaded%
    }{See the gnuplot documentation for explanation.%
    }{The gnuplot epslatex terminal needs graphicx.sty or graphics.sty.}%
    \renewcommand\includegraphics[2][]{}%
  }%
  \providecommand\rotatebox[2]{#2}%
  \@ifundefined{ifGPcolor}{%
    \newif\ifGPcolor
    \GPcolortrue
  }{}%
  \@ifundefined{ifGPblacktext}{%
    \newif\ifGPblacktext
    \GPblacktexttrue
  }{}%
  \let\gplgaddtomacro\g@addto@macro
  \gdef\gplbacktext{}%
  \gdef\gplfronttext{}%
  \makeatother
  \ifGPblacktext
    \def\colorrgb#1{}%
    \def\colorgray#1{}%
  \else
    \ifGPcolor
      \def\colorrgb#1{\color[rgb]{#1}}%
      \def\colorgray#1{\color[gray]{#1}}%
      \expandafter\def\csname LTw\endcsname{\color{white}}%
      \expandafter\def\csname LTb\endcsname{\color{black}}%
      \expandafter\def\csname LTa\endcsname{\color{black}}%
      \expandafter\def\csname LT0\endcsname{\color[rgb]{1,0,0}}%
      \expandafter\def\csname LT1\endcsname{\color[rgb]{0,1,0}}%
      \expandafter\def\csname LT2\endcsname{\color[rgb]{0,0,1}}%
      \expandafter\def\csname LT3\endcsname{\color[rgb]{1,0,1}}%
      \expandafter\def\csname LT4\endcsname{\color[rgb]{0,1,1}}%
      \expandafter\def\csname LT5\endcsname{\color[rgb]{1,1,0}}%
      \expandafter\def\csname LT6\endcsname{\color[rgb]{0,0,0}}%
      \expandafter\def\csname LT7\endcsname{\color[rgb]{1,0.3,0}}%
      \expandafter\def\csname LT8\endcsname{\color[rgb]{0.5,0.5,0.5}}%
    \else
      \def\colorrgb#1{\color{black}}%
      \def\colorgray#1{\color[gray]{#1}}%
      \expandafter\def\csname LTw\endcsname{\color{white}}%
      \expandafter\def\csname LTb\endcsname{\color{black}}%
      \expandafter\def\csname LTa\endcsname{\color{black}}%
      \expandafter\def\csname LT0\endcsname{\color{black}}%
      \expandafter\def\csname LT1\endcsname{\color{black}}%
      \expandafter\def\csname LT2\endcsname{\color{black}}%
      \expandafter\def\csname LT3\endcsname{\color{black}}%
      \expandafter\def\csname LT4\endcsname{\color{black}}%
      \expandafter\def\csname LT5\endcsname{\color{black}}%
      \expandafter\def\csname LT6\endcsname{\color{black}}%
      \expandafter\def\csname LT7\endcsname{\color{black}}%
      \expandafter\def\csname LT8\endcsname{\color{black}}%
    \fi
  \fi
  \setlength{\unitlength}{0.0500bp}%
  \begin{picture}(4680.00,3402.00)%
    \gplgaddtomacro\gplbacktext{%
      \csname LTb\endcsname%
      \put(682,704){\makebox(0,0)[r]{\strut{}$0$}}%
      \put(682,1191){\makebox(0,0)[r]{\strut{}$0.2$}}%
      \put(682,1677){\makebox(0,0)[r]{\strut{}$0.4$}}%
      \put(682,2164){\makebox(0,0)[r]{\strut{}$0.6$}}%
      \put(682,2650){\makebox(0,0)[r]{\strut{}$0.8$}}%
      \put(682,3137){\makebox(0,0)[r]{\strut{}$1$}}%
      \put(814,484){\makebox(0,0){\strut{}$0$}}%
      \put(1508,484){\makebox(0,0){\strut{}$0.2$}}%
      \put(2202,484){\makebox(0,0){\strut{}$0.4$}}%
      \put(2895,484){\makebox(0,0){\strut{}$0.6$}}%
      \put(3589,484){\makebox(0,0){\strut{}$0.8$}}%
      \put(4283,484){\makebox(0,0){\strut{}$1$}}%
      \put(176,1920){\rotatebox{-270}{\makebox(0,0){\strut{}$\Phi$}}}%
      \put(2548,154){\makebox(0,0){\strut{}$p$}}%
    }%
    \gplgaddtomacro\gplfronttext{%
      \csname LTb\endcsname%
      \put(2266,2964){\makebox(0,0)[r]{\strut{}stable}}%
      \csname LTb\endcsname%
      \put(2266,2744){\makebox(0,0)[r]{\strut{}unstable}}%
      \csname LTb\endcsname%
      \put(2266,2524){\makebox(0,0)[r]{\strut{}metastable}}%
    }%
    \gplbacktext
    \put(0,0){\includegraphics{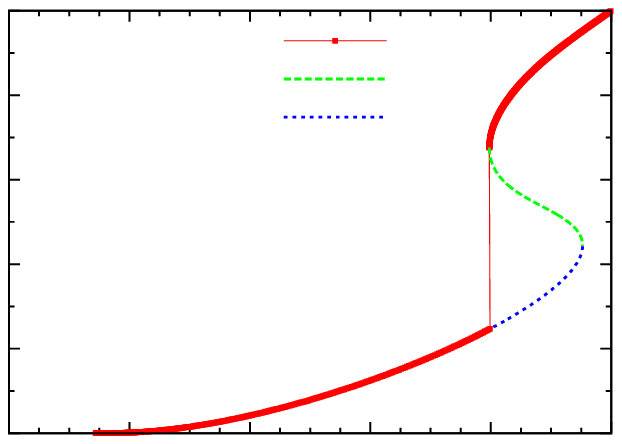}}%
    \gplfronttext
  \end{picture}%
\endgroup

%% file: preprint_JSM_2019.bbl
\section*{References}

\begin{thebibliography}{50}


\bibitem{Chalupa} J. Chalupa and P. L. Leath and G. R. Reich, {\em
  Bootstrap percolation on a Bethe lattice}, J. Phys. C: Solid State
  Phys. \textbf{12}, L31 (1979).

\bibitem{BP_rev} For a review, see: P. De Gregorio, A. Lawlor, and
  K. A. Dawson, in {\em Encyclopedia of Complexity and Systems
    Science}, edited by R.A. Meyers (New York: Springer, 2009),
  pp. 608--626.

\bibitem{Bar} A. Bar and D. Mukamel, {\em Mixed-order phase transition
  in a one-dimensional model}, Phys. Rev. Lett. {\bf 112}, 015701
  (2014); {\em Mixed order transition and condensation in an exactly
  soluble one dimensional spin model}, J. Stat. Mech. {\bf 11}, P11001
  (2014), and references therein for early examples of mixed-order
  phase transitions in one dimensional spin models with long-range
  interactions.

\bibitem{Branco} N.S. Branco, {\em Probabilistic Bootstrap
  Percolation}, J. Stat. Phys. \textbf{70}, 1035 (1993).

\bibitem{Doro} G.J. Baxter, S.N. Dorogovtsev, A.V. Goltsev, and
  J.F.F. Mendes, {\em Bootstrap percolation on complex networks},
  Phys. Rev. E \textbf{82}, 011103 (2010); {\em Heterogeneous k-core
    versus bootstrap percolation on complex networks}, Phys. Rev.  E
  \textbf{83}, 051134 (2011).

\bibitem{Cellai} D. Cellai, A. Lawlor, K.A. Dawson, and J.P. Gleeson,
  {\em Tricritical Point in Heterogeneous k-Core Percolation},
  Phys. Rev. Lett.  \textbf{107}, 175703 (2011); {\em Critical
    phenomena in heterogeneous k-core percolation}, Phys. Rev. E {\bf
    87}, 022134 (2013).

\bibitem{SeDeCaAr} M. Sellitto, D. De Martino, F. Caccioli, and
  J.J. Arenzon, {\em Dynamic Facilitation Picture of a Higher-Order
    Glass Singularity}, Phys. Rev. Lett. \textbf{105}, 265704 (2010).

\bibitem{Se2012} M. Sellitto, {\em Cooperative heterogeneous
  facilitation: Multiple glassy states and glass-glass transition},
  Phys. Rev. E \textbf{86}, 030502(R) (2012); {\em Disconnected
    glass-glass transitions and swallowtail bifurcations in
    microscopic spin models with facilitated dynamics},
  J. Chem. Phys. {\bf 138}, 224507 (2013).

\bibitem{PaSe} G. Parisi and M. Sellitto, {\em The large-connectivity
  limit of bootstrap percolation}, Europhys. Lett. {\bf 109} 36001
  (2015).

\bibitem{Dawson} K. Dawson, G. Foffi, M. Fuchs, W. G\"otze,
  F. Sciortino, M. Sperl, P.  Tartaglia, T. Voigtmann, and
  E. Zaccarelli, {\em Higher-order glass-transition singularities in
    colloidal systems with attractive interactions}, Phys. Rev. E {\bf
    63}, 011401 (2000).
 
\bibitem{Goetze} W. G\"otze, {\em Complex Dynamics of Glass-Forming
  Liquids}, (Oxford: Oxford University Press, 2009).

\bibitem{RiSo} F. Ritort, and P. Sollich, {\em Glassy dynamics of
  kinetically constrained models}, Advances in Physics \textbf{52},
  219 (2003).

\bibitem{SeBiTo} M. Sellitto, G. Biroli, and C. Toninelli, {\em
  Facilitated spin models on Bethe lattice: Bootstrap percolation,
  mode-coupling transition and glassy dynamics},
  Europhys. Lett. \textbf{69}, 496 (2005).

\bibitem{Sciortino} F. Sciortino, {\em Disordered materials: One
  liquid, two glasses}, Nature Mater. {\bf 1}, 145 (2002).

\bibitem{Pham} N. Pham, A.M. Puertas, J. Bergenholtz, S.U. Egelhaaf,
  A. Moussaid, P.N. Pusey, A.B. Schofield, M.E. Cates, M. Fuchs, and
  W.C. Poon, {\em Multiple glassy states in a simple model system},
  Science {\bf 296}, 104 (2002).

\bibitem{Voigtmann} Th. Voigtmann, {\em Multiple glasses in asymmetric
  binary hard spheres}, Europhys. Lett. {\bf 96}, 36006 (2011).

\bibitem{Gnan} N. Gnan, G. Das, M. Sperl, F. Sciortino, and
  E. Zaccarelli, {\em Multiple glass singularities and isodynamics in
    a core-softened model for glass-forming systems},
  Phys. Rev. Lett. {\bf 113}, 258302 (2014).

\bibitem{Maimbourg} T. Maimbourg, M. Sellitto, G. Semerjian, and
  F. Zamponi, {\em Generating dense packings of hard spheres by soft
    interaction design}, SciPost. Phys. {\bf 4}, 039 (2018).

\bibitem{Nelson} P. Nelson, {\em Biological Physics} (New York: WH
  Freeman, 2004).

\bibitem{Greer} A.L. Greer, {\em Too hot to melt}, Nature {\bf 404},
  134 (2000), and references therein.

\bibitem{StDe} F.H. Stillinger and P.G. Debenedetti, {\em Phase
  transitions, Kauzmann curves and inverse melting},
  Biophys. Chem. {\bf 105}, 211 (2003).

\bibitem{ScSh} N. Schupper and N.M. Shnerb, {\em Inverse melting and
  inverse freezing: A spin model}, Phys.  Rev. E {\bf 72}, 046107
(2005).

\bibitem{Plazanet} M. Plazanet et al. {\em Crystallization on heating
  and complex phase behavior of $\alpha$-cyclodextrin solutions}, J.
  Chem. Phys. {\bf 125}, 154504 (2006).

\bibitem{Schulman} L.S. Schulman and P.E. Seiden, {\em Statistical
  Mechanics of a Dynamical System based on Conway's Game of Life},
  J. Stat. Phys. {\bf 19}, 293 (1978).

\bibitem{BaReRu} F. Bagnoli, R. Rechtman, and S. Ruffo, {\em Some
  facts of life}, Physica A: Statistical Mechanics and its
  Applications {\bf 171}, 249 (1991).

\bibitem{Krako} V. Krakoviack, {\em Comment on ``Spherical 2+p
  spin-glass model: An analytically solvable model with a
  glass-to-glass transition''}, Phys Rev. B {\bf 76}, 136401 (2007).

\bibitem{CrLe} A. Crisanti and L. Leuzzi, {\em Reply to Comment on
  ``Spherical 2+p spin-glass model: an analytically solvable model
  with a glass-to-glass transition''}, Phys. Rev. B {\bf 76}, 136402
  (2007).

\bibitem{Rizzo} T. Rizzo, {\em Fate of the Hybrid Transition of
  Bootstrap Percolation in Physical Dimension},
  Phys. Rev. Lett. \textbf{122}, 108301 (2019).

\bibitem{RiVo} T. Rizzo and Th. Voigtmann, {\em Solvable Models of
  Supercooled Liquids at the Avoided Mode-Coupling-Theory Transition},
  arXiv preprint arXiv:1903.01773 (2019).

\bibitem{PaRi} G. Parisi and T. Rizzo, {\em k-core percolation in four
  dimensions}, Phys. Rev. E \textbf{78}, 022101 (2008).


\end{thebibliography}
